\documentclass[pdflatex,sn-mathphys-num]{sn-jnl}

\usepackage{amsmath,amssymb,amsfonts} 
\usepackage{amsthm} 
\usepackage{mathrsfs} 
\usepackage{array} 
\usepackage{booktabs} 
\usepackage{adjustbox} 

\usepackage{graphicx} 
\usepackage{subcaption} 
\usepackage{tikz} 

\usepackage{manyfoot} 
\usepackage{enumitem} 
\usepackage{setspace} 

\usepackage{algorithm}
\usepackage{algorithmicx}
\usepackage{algpseudocode}

\usepackage{xcolor}

\theoremstyle{thmstyleone}%
\theoremstyle{thmstyletwo}%

\theoremstyle{thmstylethree}%

\begin{document}

\title[Article Title]{FMD-TransUNet: Abdominal Multi-Organ Segmentation Based on Frequency Domain Multi-Axis Representation Learning and Dual Attention Mechanisms}


\author[1]{\fnm{Fang} \sur{Lu}}

\author[1]{\fnm{Jingyu} \sur{Xu}}

\author[1]{\fnm{Qinxiu} \sur{Sun}}

\author*[1]{\fnm{Qiong} \sur{Lou}} \email{bearqiong@163.com}

\affil[1]{%
\orgdiv{School of Science},%
\orgname{ Zhejiang University of Science and Technology},%
\orgaddress{%
\street{ No.318 Liuhe Road},%
\city{ Hangzhou},%
\postcode{ 310023},%
\state{ Zhejiang},%
\country{ China}%
}%
}





\abstract{Accurate abdominal multi-organ segmentation is critical for clinical applications. Although numerous deep learning-based automatic segmentation methods have been developed, they still struggle to segment small, irregular, or anatomically complex organs. Moreover, most current methods focus on spatial-domain analysis, often overlooking the synergistic potential of frequency-domain representations. To address these limitations, we propose a novel framework named FMD-TransUNet for precise abdominal multi-organ segmentation. It innovatively integrates the Multi-axis External Weight Block (MEWB) and the improved dual attention module (DA+) into the TransUNet framework. The MEWB extracts multi-axis frequency-domain features to capture both global anatomical structures and local boundary details, providing complementary information to spatial-domain representations. The DA+ block utilizes depthwise separable convolutions and incorporates spatial and channel attention mechanisms to enhance feature fusion, reduce redundant information, and narrow the semantic gap between the encoder and decoder. Experimental validation on the Synapse dataset shows that FMD-TransUNet outperforms other recent state-of-the-art methods, achieving an average DSC of 81.32\% and a HD of 16.35 mm across eight abdominal organs. Compared to the baseline model, the average DSC increased by 3.84\%, and the average HD decreased by 15.34 mm. These results demonstrate the effectiveness of FMD-TransUNet in improving the accuracy of abdominal multi-organ segmentation.}




\maketitle

\section{Introduction}

\begin{figure}[ht]
   \begin{center}
   \includegraphics[width=\textwidth, trim=0cm 5.6cm 0cm 6cm, clip]{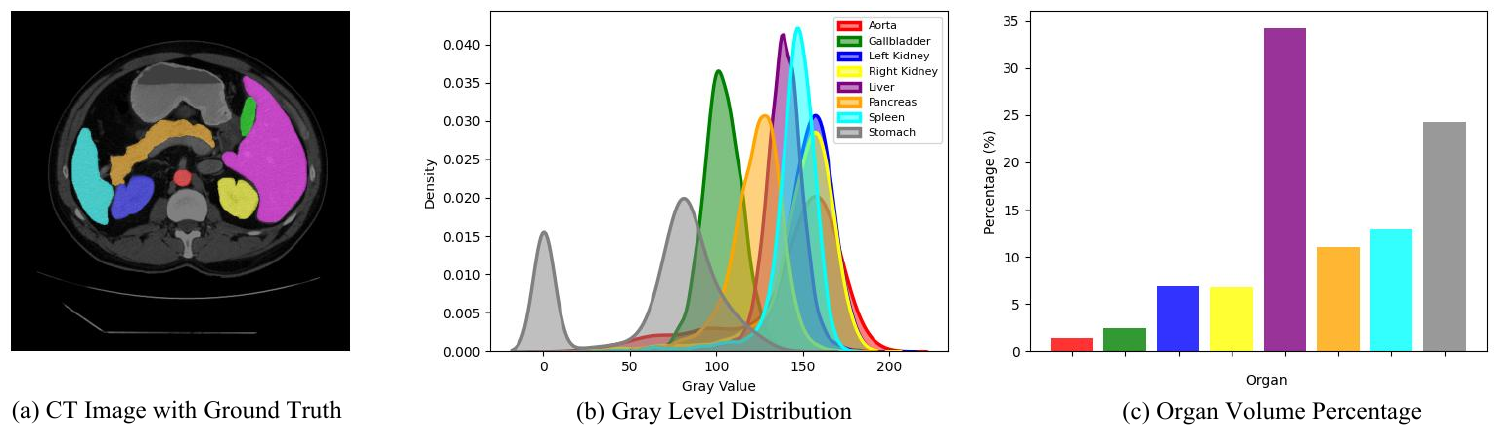} 
   %
   %
   \caption{Challenges in accurate abdominal multi-organ segmentation. (a) CT Image with Ground Truth, (b) Gray Level Distribution, (c) Organ Volume Percentage}
   \label{f:1}
   \end{center}
\end{figure}

Accurate multi-organ segmentation in computed tomography (CT) imaging is essential for a wide range of clinical applications such as disease diagnosis, radiotherapy planning, and surgical navigation \cite{ref1}. However, manual annotation by radiologists is time-consuming and subject to significant inter-observer variability, especially when applied to large-scale datasets. As illustrated in Fig.~\ref{f:1}, this task presents several major challenges. First, abdominal CT images exhibit complex organ shapes and ambiguous boundaries (Fig.~\ref{f:1} (a)). Second, overlapping grayscale intensity distributions—particularly among the liver, spleen, and kidneys—lead to difficulties in differentiating adjacent structures (Fig.~\ref{f:1} (b)). Third, the size disparity among organs is significant. For example, the liver can occupy more than 30\% of the abdominal cavity, while smaller organs such as the gallbladder and pancreas represent less than 5\% (Fig.~\ref{f:1} (c)). These factors highlight the urgent need for robust and automated segmentation tools to improve both diagnostic consistency and clinical efficiency.

In recent years, deep learning has significantly advanced automated multi-organ segmentation. The U-Net architecture, with its U-shaped encoder–decoder structure, has become a cornerstone for medical image segmentation tasks \cite{ref2}. To address specific limitations of the original U-Net, several variants have been developed, including U-Net++ \cite{ref3}, Attention U-Net \cite{ref4}, and Residual U-Net \cite{ref5}. Further extensions such as 3D U-Net \cite{ref6}, Dense U-Net \cite{ref7}, and DAResUNet \cite{ref8} have introduced volumetric representations and attention mechanisms to improve performance. Although these convolutional models are effective at capturing local spatial features, their limited receptive fields restrict the ability to model long-range anatomical dependencies—an essential aspect for accurately segmenting overlapping and morphologically complex organs.

To overcome this limitation, Transformer-based models have gained attention due to their superior capacity to model global context and long-range dependencies. Initially introduced in natural language processing, Transformers have shown strong potential in vision tasks through their self-attention mechanisms \cite{ref9}. In the domain of medical imaging, Vision Transformers (ViTs) \cite{ref10} have been successfully adapted to capture rich global semantic information. Hybrid models such as TransUNet \cite{ref11} and UCTransNet \cite{ref12} integrate Transformer encoders into the U-Net framework, combining the strengths of both architectures—preserving local feature extraction capabilities while incorporating global context modeling. For instance, Swin-Unet \cite{ref13} replaces convolutional layers with Swin Transformer blocks, enabling hierarchical representations for better segmentation performance. DA-TransUNet \cite{ref14} further enhances this structure by embedding dual attention mechanisms to extract both spatial and channel-dependent features, improving segmentation of smaller and more complex organs. Nevertheless, the Transformer lacks intrinsic mechanisms to consider image-specific attributes such as spatial position and channel information. This limitation can reduce their ability to distinguish fine anatomical structures, especially when organ boundaries are subtle or overlapping. Further investigation is needed to address this gap.

Moreover, most current segmentation methods focus on spatial-domain, which are effective for capturing local textures and edges. However, they often struggle with organs that have uneven gray-level distributions, irregular shapes, or ambiguous boundaries, which are common problems in abdominal imaging. When spatial-domain methods face limitations in feature extraction or performance improvement, frequency-domain approaches offer a complementary representation. These methods can reveal differences in signal strength that are unobtainable in the spatial domain, providing new perspectives for investigating complex anatomical structures and enhancing segmentation accuracy. Recent models such as GFNet \cite{ref15} and GFUNet \cite{ref16} have improved segmentation accuracy by leveraging Fourier transforms to disentangle foreground-background features. However, they rely on uniaxial frequency representations, which are not sufficient for modeling the anatomical complexity in multi-organ segmentation. In contrast, multiaxial frequency analysis captures directional features across multiple spatial dimensions. This offers richer and more discriminative representations of both global structures and local details. MEW-UNet \cite{ref17} exemplifies this approach by incorporating a Multi-axis External Weights (MEW) module to enhance tissue discrimination. Despite their potential, frequency-domain methods remain underexplored in abdominal multi-organ segmentation, particularly in addressing challenges posed by significant organ size discrepancies, boundary ambiguities, and heterogeneous intensity distributions, where frequency analysis could provide discriminative features through multi-axis frequency decomposition.

To mitigate the aforementioned limitations and further improve the accuracy of abdominal multi-organ segmentation, we propose a novel framework called FMD-TransUNet. This model integrates a Multi-axis External Weight Block (MEWB) and an enhanced Dual Attention (DA+) module into the TransUNet framework to jointly leverage spatial-domain and frequency-domain representations. Specifically, the MEWB applies multi-axis two-dimensional Discrete Fourier Transforms (DFT) to extract directional frequency features across spatial dimensions. These features complement spatial-domain information by providing global context and enhancing boundary discrimination, which is particularly beneficial for small or complex organs such as such as the gallbladder and kidneys. Meanwhile, the DA+ module replaces standard convolutions with depthwise separable convolutions and combines spatial and channel attention mechanisms. This design improves the extraction of semantic and regional features with reduced computational cost, which is especially effective for segmenting larger organs like the liver and spleen. By embedding the DA+ module both before the Transformer layers and within skip connections, the model further strengthens multi-scale feature integration and mitigates the semantic gap between encoder and decoder, thereby improving segmentation accuracy across organs with diverse anatomical characteristics.

The main contributions of our work are summarized as follows:

1.We propose a novel transformer-based hybrid architecture called FMD-TransUNet, integrating multi-axis frequency-domain features and enhanced attention mechanisms to  improve multi-organ segmentation in abdominal CT images.

2.We introduce the MEWB module to extract global-local frequency features that complement spatial representations, and the DA+ module to enhance feature fusion and reduce the semantic gap between the encoder and decoder.

3.We conduct extensive evaluations of FMD-TransUNet on the Synapse dataset, and it outperforms the state-of-the-art models, particularly in segmenting anatomically complex organs such as the pancreas and kidneys.

\section{Related Work}

\subsection{Transformer}

The Transformer model has gained considerable attention in the field of medical image segmentation, attributed to its ability to capture long-range dependencies. The Vision Transformer (ViT) \cite{ref18} was initially introduced for image classification and later extended to other tasks, such as motion detection \cite{ref19} and image segmentation \cite{ref20}. TransUNet \cite{ref11} has seamlessly integrated the U-Net architecture with Transformer technology, thereby enhancing the accuracy of multi-organ segmentation tasks. Subsequently, models such as Swin-Unet \cite{ref13} and nnFormer \cite{ref21} further refined the approach. These models have replaced traditional convolutional blocks with more efficient Transformers or integrated convolution with self-attention to improve feature extraction. TransNorm \cite{ref22} introduced normalization techniques to improve feature alignment and stability in multi-organ segmentation. However, an increase in the number of Transformer layers in these models results in higher computational costs. To address this problem, CTC-Net \cite{ref23} and HiFormer \cite{ref24} proposed combining CNNs with Transformers to extract complementary features and improve fusion through attention mechanisms. Despite these improvements, many models are still prone to overfitting due to their complexity. Additionally, the Transformer lacks intrinsic mechanisms to account for image-specific attributes, such as spatial position and channel information. Inspired by the success of TransUNet in abdominal multi-organ segmentation, our method incorporates frequency-domain information and dual-attention mechanisms into TransUNet. This approach aims to enhance feature extraction for organs with irregular shapes, uneven gray levels, and unclear boundaries.

\subsection{Frequency domain representation learning}

Frequency domain methods have shown potential in improving medical image segmentation, yet they remain relatively underexplored compared to spatial domain approaches. Hybrid models that combine Fourier transforms with CNNs or Transformers have been proposed to leverage frequency-domain features and improve segmentation accuracy \cite{ref25}. For instance, FDFUNet \cite{ref26} is a multi-scale frequency-domain filtering network. It uses the multi-scale frequency domain filter to extract global frequency features and fuses them with spatial-domain features to improve segmentation accuracy. FSSN \cite{ref27} is a frequency selection network. It suppresses irrelevant frequency components to reduce semantic gaps, achieving higher accuracy with fewer parameters and lower computational complexity. The frequency-enhanced lightweight vision Mamba network \cite{ref28} is another example that explores lightweight designs while maintaining segmentation accuracy. These studies show that frequency-domain methods can analyze the frequency components of an image, which helps in identifying patterns and structures that are not easily distinguishable using only spatial information. This enables superior feature extraction. However, most existing methods only focus on single-axis frequency information, limiting their ability to distinguish complex signals. To overcome this limitation, we introduce the MEWB into the TransUNet framework. This block is designed to capture multi-axis frequency features, enhancing both global-local feature representation and complementing spatial-domain information. 

\subsection{Dual attention mechanism}

Dual attention mechanisms are widely used in medical image segmentation due to their ability to capture spatial and channel-wise dependencies. Early works, such as DANet \cite{ref29}, introduced parallel spatial and channel attention to model long-range dependencies, substantially improving segmentation accuracy in scene segmentation tasks. This approach was subsequently extended to medical image analysis. Specifically, the DoubleU-Net \cite{ref30} integrated dual attention mechanisms within the U-Net architecture and achieved superior performance in the detection of small and irregular structures, such as polyps. Similarly, a dual attention encoder-decoder network has been proposed, effectively fusing spatial and channel features to address segmentation challenges in complex anatomical patterns \cite{ref31}. Recent developments have incorporated dual attention mechanisms into Transformer-based architectures. DaViT \cite{ref32} sequentially applied spatial and channel attention to progressively enhance feature representations. Channel cross-attention has been introduced into U-Net to reduce the semantic gap between encoder and decoder features, thereby improving segmentation accuracy on challenging datasets \cite{ref33}. DA-TransUNet \cite{ref14} further integrates dual attention mechanisms to enhance spatial and channel-specific feature extraction for medical image segmentation.

However, dual attention mechanisms and Transformer-based models often encounter limitations, including feature redundancy, inefficient feature extraction, and the neglect of image-specific attributes. To address these challenges, we replaced the standard convolutions in the dual attention mechanism with depthwise separable convolutions. This modification reduces redundant information and narrows the semantic gap while maintaining model performance. In addition, the integration of the MEWB and DA+ blocks enables more efficient feature fusion, making it more effective for organs with irregular shapes or ambiguous boundaries, and further improving the model’s segmentation performance.

\begin{figure}[ht]
   \begin{center}
   \includegraphics[width=0.85\textwidth, height=0.35\textheight, trim=0cm 1.85cm 0cm 1cm, clip]{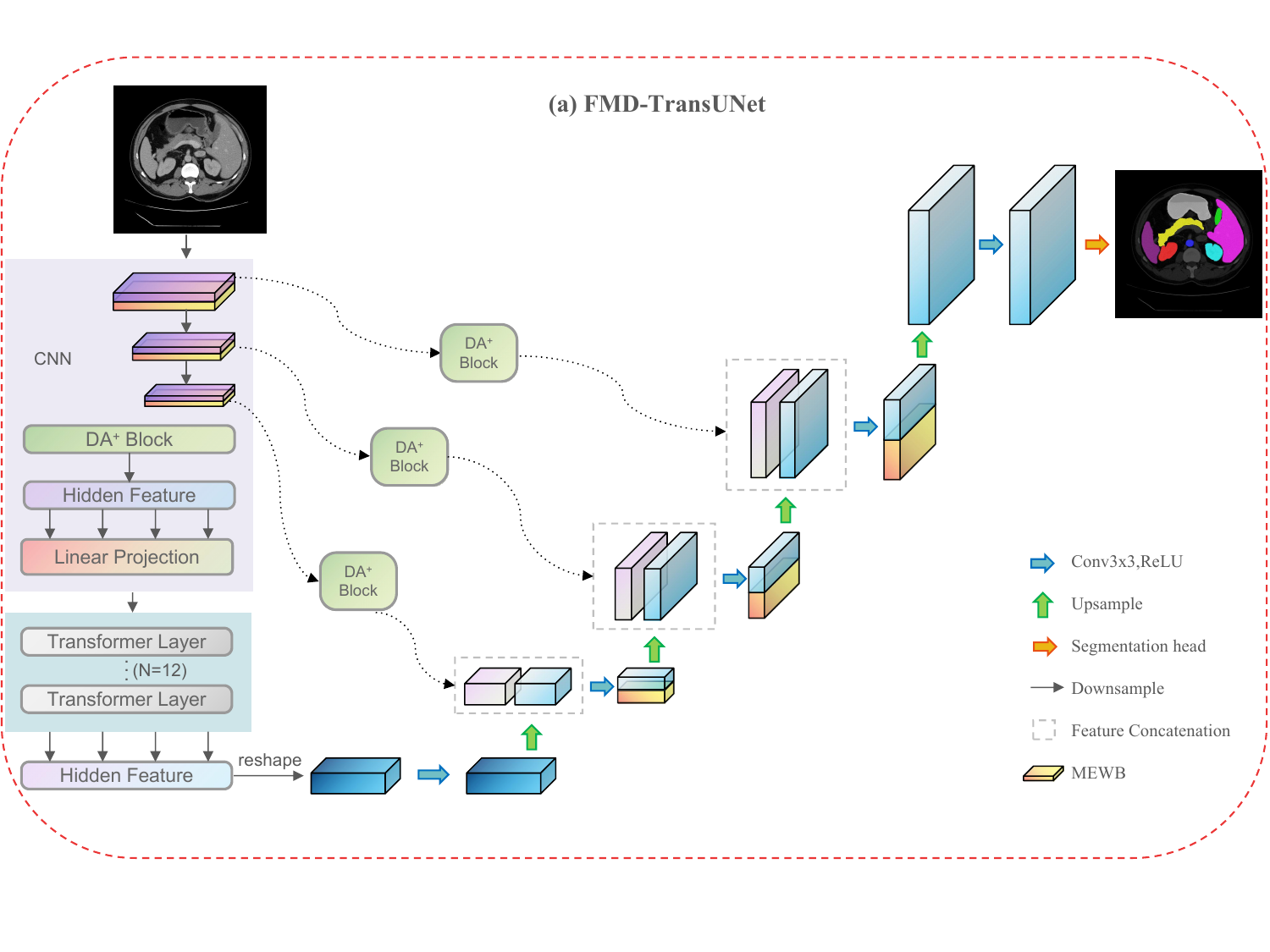} 
   \includegraphics[width=0.85\textwidth, height=0.25\textheight, trim=0cm 7.3cm 0cm 0.1cm, clip]{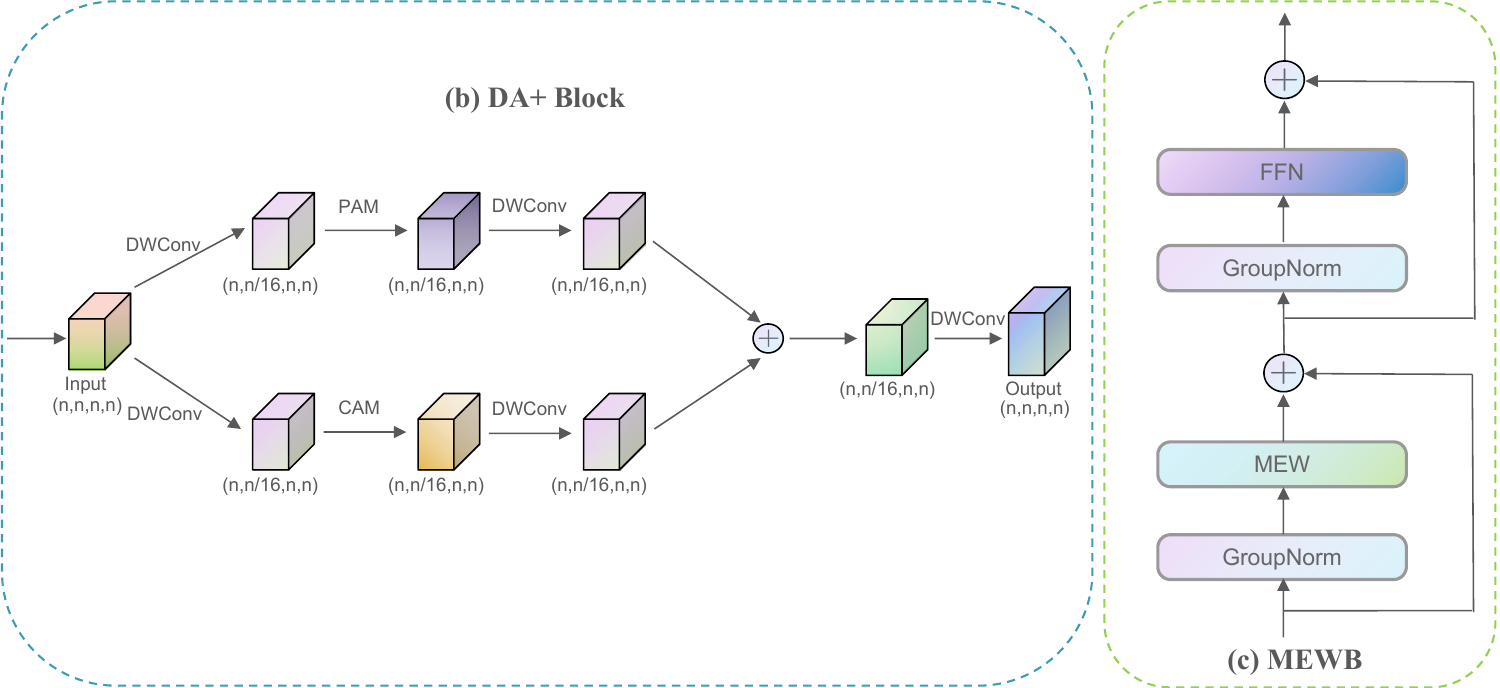}
   %
   %
   \caption{(a) Architecture of the proposed FMD-TransUNet. (b) Schematic of DA+ Block. The same input feature map is input into two feature extraction layers, one is the position feature extraction block and the other is the channel feature extraction block. (c) Schematic of MEWB. FFN presents the feed-forward layer.}
   \label{f:2}
   \end{center}
\end{figure}

\section{Method}

In this section, we will present the basic structure and overall process of FMD-TransUNet. The MEWB and DA+ blocks used in the encoder, decoder and skip connections will be explained respectively.

\subsection{Overview}

As shown in Fig.~\ref{f:2}, the proposed FMD-TransUnet comprises three main components: the encoder, the decoder, and skip connections. The encoder combines convolutional neural networks with Transformer layers, strategically integrating the MEWB and DA+ blocks to enhance feature extraction. The decoder reconstructs feature maps using traditional convolution operations combined with MEWB modules, ensuring effective recovery of spatial details. To optimize skip connections, DA+ blocks act as a key component, refining transmitted features by filtering out irrelevant information and bridging the semantic gap between the encoder and decoder. Compared to traditional convolutional methods and Transformer-heavy models, FMD-TransUNet specially integrates both multi-axis frequency domain information and dual attention mechanisms by MEWB and DA+ block respectively. This integration enhances the model's ability to capture both global and local features, improving segmentation accuracy and robustness. Additionally, it helps in handling the variability in organ shapes and sizes, and reduces the interference from complex backgrounds, which is crucial for accurate multi-organ segmentation in abdominal medical images. 

The encoder consists of five key components: the convolution block, the MEWB module, the DA+ block, the embedding layer and the Transformer layer. The encoder begins with three convolutional blocks that progressively downsample the input feature maps while increasing the number of channels. These operations extract low-level spatial features and expand the receptive field, ensuring efficient feature representation. After the convolutional layers, the MEWB module is applied to enhance feature maps by leveraging multi-axis frequency domain information and global-local representations. Following MEWB, the DA+ block is used to refine features before passing them to Transformer layers. The Transformer layers capture long-range dependencies and global context, enabling a comprehensive understanding of the image structure.

The skip connections bridge the encoder and decoder, transmitting features directly between corresponding layers. To reduce the semantic gap and eliminate redundancy in traditional skip connections, DA+ blocks are introduced within the skip connections. These blocks refine the transmitted features by focusing on spatial and channel-specific dependencies, ensuring the decoder receives high-quality feature maps. This refinement improves the reconstruction of details and segmentation precision, especially for small or complex anatomical structures.

The decoder in FMD-TransUNet complements the encoder by progressively reconstructing feature maps to their original resolution. It consists of three upsampling blocks, feature fusion, and a final segmentation head. Each upsampling block is followed by the MEWB module, which refines the features at every stage. During each upsampling operation, the decoder fuses the current feature maps with refined features transmitted through the DA+ block-enhanced skip connections. This integration effectively combines the global context captured by the encoder and Transformer layers with the local details preserved in the skip connections, thereby enhancing the feature representation. The MEWB module is applied after each upsampling step to further refine the feature maps by integrating global and local information, ensuring precise segmentation of complex structures. Finally, the segmentation head restores the feature maps to their original resolution, producing the final segmentation map with pixel-level accuracy. This architecture enables the decoder to effectively utilize encoder and skip connection features, resulting in robust and accurate segmentation. 

\subsection{The Multi-axis External Weights Block}

Current medical image segmentation approaches predominantly focus on spatial-domain analysis, often overlooking the potential of frequency-domain representations. While spatial methods struggle to delineate ambiguous boundaries due to noise or low contrast, frequency-domain analysis enables precise extraction of structural features by isolating critical frequency components corresponding to edges and textures, thereby offering a complementary pathway to enhance segmentation accuracy. However, single-axis frequency analysis remains limited due to its orientation-constrained sensitivity. To address this limitation, we introduce a multi-axis frequency domain module (MEWB) between the corresponding layers of the encoder and decoder, enabling robust and comprehensive boundary characterization across diverse anatomical geometries. Its structure is shown in Fig.~\ref{f:2} (c).

\begin{figure}[ht]
   \begin{center}
   \includegraphics[width=\textwidth, trim=4cm 7.1cm 5cm 4.9cm, clip]{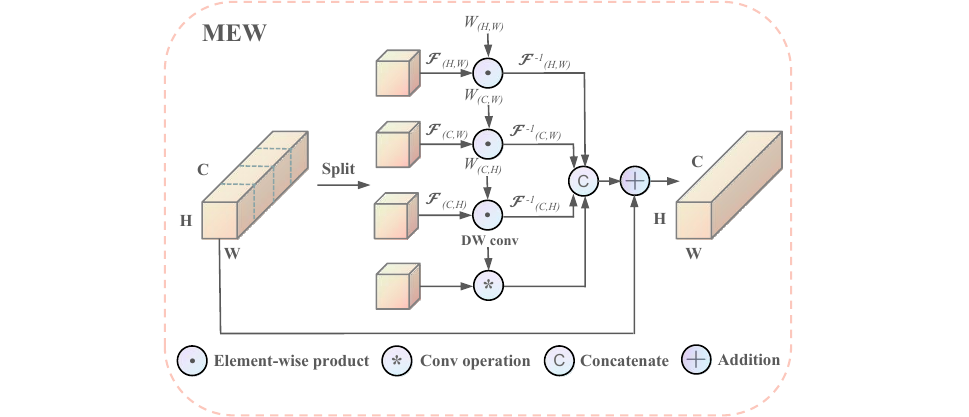} 
   %
   %
  \caption{The Multi-axis External Weights (MEW) mechanism. 
\(\mathcal{F}_{(H, W)}\), \(\mathcal{F}_{(C, W)}\) and \(\mathcal{F}_{(C, H)}\) refer to conducting 2D DFT along the height-width, channel-width, and channel-height axes respectively. \(\mathcal{F}^{-1}_{(H, W)}\), \(\mathcal{F}^{-1}_{(C, W)}\), and \(\mathcal{F}^{-1}_{(C, H)}\) are corresponding 2D inverse DFT. FFN presents the feed-forward layer.}
   \label{f:3} 
   \end{center}
\end{figure}

First, the input feature maps are normalized using GroupNorm \cite{ref34} to stabilize training and improve convergence. Then, the MEW module processes the feature maps to extract global-local information in the frequency domain. The MEW module consists of four branches, as shown in Fig.~\ref{f:3}. The first three branches focus on extracting global information in the multi-axis frequency domain, while the fourth branch captures local information in the spatial domain to enhance feature representation. Specifically, the MEW module applies 2D DFT along three different axes: height-width, channel-width, and channel-height. For each axis, learnable external weights are used to adjust the frequency domain features. The adjusted features are then transformed back to the spatial domain using the Inverse 2D DFT. In the fourth branch of the MEW module, Depthwise Separable Convolution (DWConv) \cite{ref40} is applied to extract local information. The outputs from all four branches are concatenated along the channel dimension to form a unified feature map. Finally, the FFN processes the fused features, and residual connections are applied at two points in the block to preserve the original information. This design ensures that MEWB captures both the global information from the frequency domain and the local information from the spatial domain, enhancing the segmentation model's ability to differentiate complex structures. 

\subsection{The improved dual attention block}

Transformer-based models often suffer from feature redundancy, inefficient local-global integration, and high computational costs. To address these issues and improve abdominal multi-organ segmentation accuracy, we propose the enhanced dual attention (DA+) block. This module synergizes spatial and channel attention mechanisms through DWConv for efficient feature refinement. As illustrated in Fig.~\ref{f:2} (b), the DA+ block comprises two parallel branches named the Position Attention Module (PAM) and the Channel Attention Module (CAM), which are adapted from the Dual Attention Network for scene segmentation \cite{ref36}. 
 
The PAM captures long-range spatial dependencies by modeling position relationships in the input feature map to enhance global contextual awareness, while the CAM models inter-channel correlations to amplify diagnostically critical features. Both branches employ DWConv to compress channel dimensions, reducing parameters and computational complexity without sacrificing performance \cite{ref35}. The refined spatial and channel attention maps are aggregated and restored to the original channel dimensions through a final DWConv layer. This process can be represented by the following equations:

\begin{equation}
\eta_{1} = DWConv(PAM(DWConv(Input)))
\label{eq:1}
\end{equation}
\begin{equation}
\eta_{2} = DWConv(CAM(DWConv(Input)))
\label{eq:2}
\end{equation}
\begin{equation}
F_{\text{output}} = DWConv(\eta_{1} + \eta_{2})
\label{eq:3}
\end{equation}

where \textit{DWConv(.)} refers to the depthwise separable convolution process, \textit{PAM(.)} and \textit{CAM(.)} denote the operation of capturing spatial dependencies and extracting channel features respectively.

Strategically integrated before Transformer layers and within encoder-decoder skip connections, the DA+ Block serves dual roles: (1) refining input features for enhanced global modeling in Transformers, and (2) filtering redundant features in skip connections to bridge semantic gaps. 

\subsection{Skip connections with the DA+ block}

Skip connections bridge the encoder and decoder to mitigate semantic gaps by directly transmitting hierarchical features. However, traditional skip connections often propagate redundant or noisy features, which may degrade segmentation accuracy \cite{ref37}. To address this limitation, we integrate DA+ blocks into three skip connection layers as shown in Fig. ~\ref{f:2} (a). 

The DA+ block employs depthwise separable convolutions within their dual attention pathways to dynamically recalibrate spatial relationships through the PAM and optimize channel-wise dependencies via the CAM. This dual refinement ensures that transmitted features retain both global anatomical context and localized boundary details. By suppressing the redundant spatial-channel information, the refined features enable the decoder to reconstruct segmentation maps with improved fidelity. Quantitative ablation studies (as shown in Section 4.3) confirm that progressively integrating DA+ blocks across multiple skip connection layers yields cumulative performance gains, achieving optimal results when applied to all three layers.

\section{Experimental results}

\subsection{Datasets and evaluation}
Synapse \cite{ref38} is a public abdominal multi-organ dataset, which contains 30 CT scans with a total of 3779 axial CT images. These images cover eight abdominal organs including aorta, gallbladder, spleen, left kidney, right kidney, liver, pancreas, and stomach. The images were randomly divided into 2212 axial slices and 1567 axial slices for training and testing respectively.

To evaluate segmentation performance, we employ two complementary metrics: the Dice Similarity Coefficient (DSC) for volumetric overlap analysis and the Hausdorff Distance for boundary alignment assessment.
The DSC quantifies the spatial agreement between predicted (P) and ground truth (T) segmentation masks as the following equation,

\begin{equation}
\text{DSC} = \frac{2 |P \cap T|}{|P| + |T|} 
\label{eq:6}
\end{equation}

DSC values range from 0 (no overlap) to 1 (perfect alignment), with higher values indicating superior segmentation accuracy.
For boundary precision evaluation, we compute the HD as follows,

\begin{equation}
\text{HD} = \max \left\{ \max_{a \in A} \min_{b \in B} \|a - b\|, \max_{b \in B} \min_{a \in A} \|b - a\| \right\} 
\label{eq:7}
\end{equation}
where A and B represent the boundary point sets of the predicted and ground truth respectively, $\|a - b\|$ represents the Euclidean distance between points $a$ and $b$. A lower HD value indicates better boundary alignment.

\subsection{Implementation details}

We implemented our method in Pytorch using a single NVIDIA\textsuperscript{\textregistered} GeForce RTX\textsuperscript{\texttrademark} 4090 D (24GB) GPU. The deep neural network was based on TransUNet, a commoly used architecture for medical image segmentation. In the data preprocessing stage, several augmentation techniques were applied to alleviate overfitting and enhance the model's robustness. Random rotation and flipping were performed to introduce variability in the training data. Additionally, random noise and random contrast adjustments were added to further enrich the dataset.

For the model architecture, the pre-trained R50-ViT model \cite{ref39} was utilized. The input images were resized to a uniform dimension of 224$\times$224 pixels, with a patch size $P$ set to 16. The training process was configured with a batch size of 6 to balance computational efficiency and model performance. The loss function was a combination of cross-entropy loss and Dice loss, which effectively balanced classification accuracy and segmentation overlap. The model was trained using the SGD optimiser with a learning rate of 0.01, momentum of 0.9, weight decay of $1 \times 10^{-4}$, and a maximum number of training sessions of 200.

\subsection{Comparison with State-of-the-arts}

Experiments were conducted on the Synapse dataset to evaluate the performance of FMD-TransUNet in comparison with several state-of-the-art models, including U-Net \cite{ref2}, U-Net++ \cite{ref3}, Residual U-Net \cite{ref5}, AttnUNet \cite{ref4}, TransUNet \cite{ref11}, TransNorm \cite{ref22}, Swin-Unet \cite{ref13}, DA-TransUNet \cite{ref14}, and MEW-UNet \cite{ref17}. The detailed results are presented in Table~\ref{tab:1}.

\begin{table*}[htbp]

\begin{center}
\caption{ Comparative experimental results on the Synapse dataset}
\label{tab:1} 
\adjustbox{max width=\textwidth}{ 
\begin{tabular}{lccccccccccc}
\hline
Model               & Year & DSC(\%)$\uparrow$ & HD(mm)$\downarrow$ & Aorta  & Gallbladder & Kidney (L) & Kidney (R) & Liver  & Pancreas & Spleen & Stomach \\
\hline
U-net \cite{ref2}      & 2015 & 76.85         & 39.70          & 89.07  & \textbf{69.72}  & 77.77      & 68.60      & 93.43  & 53.98    & 86.67  & 75.58   \\
U-Net++ \cite{ref3}    & 2018 & 76.91         & 36.93          & 88.19  & 68.89          & 81.76      & 75.27      & 93.01  & 58.20    & 83.44  & 70.52   \\
Residual U-Net \cite{ref5} & 2018 & 
76.95     & 38.44          & 87.06  & 66.05          & 83.43      & 76.83      & 93.99  & 51.86    & 85.25  & 70.13   \\
AttnU-Net \cite{ref4}    & 2018 & 77.77         & 36.02          & \textbf{89.55}  & 68.88          & 77.98      & 71.11      & 93.57  & 58.04    & 87.30  & 75.75   \\
TransUNet \cite{ref11}           & 2021 & 77.48        & 31.69          & 87.23  & 63.13          & 81.87      & 77.02      & 94.08  & 55.86    & 85.08  & 75.62   \\
TransNorm \cite{ref22}           & 2022 & 78.40         & 30.25          & 86.23  & 65.10          & 82.18      & 78.63      & 94.22  & 55.34    & 89.50  & 76.01   \\
Swin-Unet \cite{ref13}           & 2022 & 79.13         & 21.55          & 85.47  & 66.53          & 83.28      & 79.61      & 94.29  & 56.58    & \textbf{90.66}  & 76.60   \\
DA-TransUNet \cite{ref14}        & 2024 & 79.80         & 23.48          & 86.54  & 65.27          & 81.70      & 80.45      & \textbf{94.57}  & 61.62    & 88.53  & \textbf{79.73}   \\
MEW-UNet \cite{ref17}            & 2024 & 78.92        & 16.44          & 86.68  & 65.32          & 82.87      & 80.02      & 93.63  & 58.36    & 90.19  & 74.26   \\
FMD-TransUNet & 2024 & \textbf{81.32} & \textbf{16.35} & 88.76  & 65.23          & \textbf{85.12} & \textbf{82.12} & 94.19  & \textbf{66.31}    & 89.73  & 79.13   \\
\hline
\end{tabular}
} 
\end{center}
 \vspace{-2em} 
\begin{flushleft}
\fontsize{9}{10}\selectfont {The bold font indicates the best}
\end{flushleft}
\end{table*}

The proposed FMD-TansUNnet achieves the best performance across eight abdominal organs on average. Specifically, it attains a highest mean DSC of 81.32\% and a lowest HD of 16.35 mm, surpassing all comparative methods. Notably, the model exhibits exceptional accuracy in segmenting anatomically complex and small organs, where conventional models often struggle. For instance, in pancreas segmentation, a challenging task due to the organ’s irregular morphology and low contrast, the model achieves a DSC of 66.31\%, surpassing TransUNet by 10.45\% and DA-TransUNet by 4.69\%. Similarly, for the left and right kidneys, which frequently exhibit overlapping intensity distributions with adjacent tissues, FMD-TransUNet attains DSC values of 85.12\% and 82.12\%, respectively, outperforming TransUNet by 3.25\% and 5.10\%. These improvements are attributed to MEWB, which leverages frequency components to resolve ambiguous boundaries, and the DA+ module, which refines spatial-channel dependencies through computationally efficient depthwise convolutions. DA-TransUNet shows slightly superior performance in liver segmentation and stomach segmentation in terms of DSC, surpassing our model by 0.38\% and 0.60\%, respectively. However, FMD-TransUNet demonstrates more robustness in boundary precision, as evidenced by its superior average HD of 16.35 mm across all eight abdominal organs. This represents a 48.4\% improvement over TransUNet (31.69 mm) and a 7.13 mm reduction compared to DA-TransUNet (23.48 mm). The significant reduction in HD highlights the model’s ability to achieve balanced performance in both regional accuracy and boundary alignment, a critical requirement for clinical applications.

\begin{figure}[ht]
   \begin{center}
   \includegraphics[width=\textwidth, trim=2cm 2.2cm 1.5cm 0cm, clip]{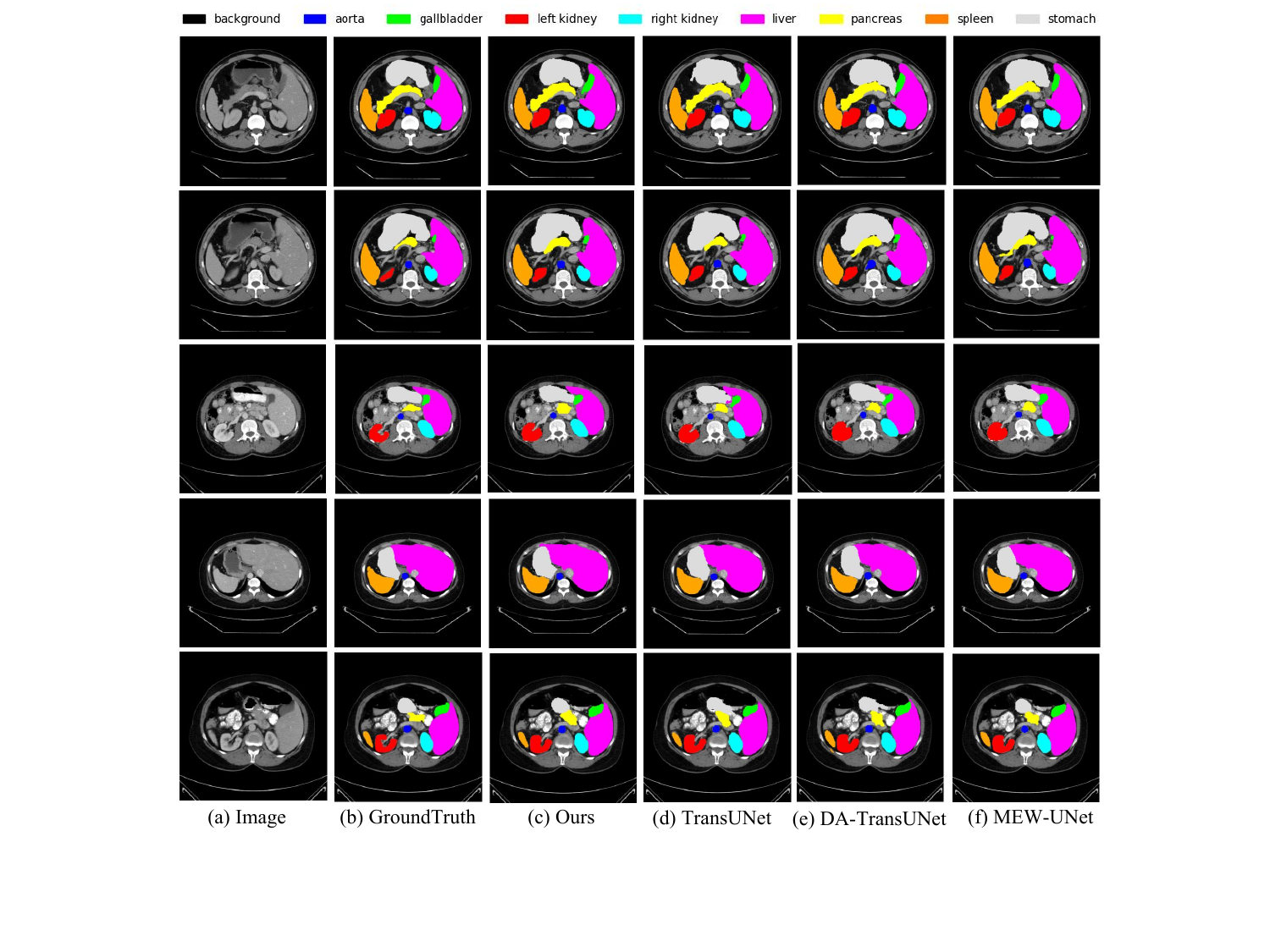} 
   %
   %
   \caption{Qualitative visualization of FMD-TransUNet segmentation results compared to other models. From left to right: (a) Image, (b) GroundTruth, (c) Ours, (d) TransUNet, (e) DA-TransUNet, (f) MEW-UNet}
   \label{f:4} 
   \end{center}
\end{figure}

The qualitative results in Fig.~\ref{f:4} further validate these quantitative findings. Compared to methods like TransUNet, DA-TransUNet, and MEW-UNet, ours more closely aligns with the ground truth segmentation. It can be clearly observed that FMD-TransUNet is capable of accurately segmenting organs of diverse sizes, ranging from large ones like the liver to small ones such as the aorta and pancreas. It also performs well in handling organs with various shapes, whether they are regular (such as the liver and gallbladder) or irregular (such as the pancreas and kidneys). Additionally, the model effectively handles organs with uniform gray levels (such as the spleen and liver) as well as those with non-uniform gray levels (such as the stomach). The high agreement with the standard segmentation results indicates that the proposed model has certain advantages in feature extraction.

In summary, FMD-TransUNet addresses the challenges of varying shapes, sizes, and boundary ambiguity through complementary frequency-domain and attention-driven feature refinement. The consistency between the visual and numerical results firmly validates the effectiveness and superiority of our approach in multi-organ segmentation.

\section{Disscusion}

To validate the contributions of each component in FMD-TransUNet, we conducted ablation studies on the Synapse dataset, focusing on three aspects: loss function weighting, module efficacy (MEWB and DA+), and DA+ block placement in skip connections.

\subsection{Impact of loss function weighting}

In our model, the loss function is a composite of Cross-Entropy Loss (\( L_c \)) and Dice Loss (\( L_d \)), as illustrated by Eq.~(\ref{eq:8}).

\begin{equation}
\text{Loss} = w_{c} \times \text{\( L_c \)} + w_{d} \times \text{\( L_d \)}
\label{eq:8}
\end{equation}
where \( w_d \) and \( w_c \) are the weights of \( L_c \) and \( L_d \), respectively. \( L_c \) and \( L_d \) are calculated according to the following Eq.~(\ref{eq:9}) and Eq.~(\ref{eq:10}), respectively.

\begin{equation}
\text{\( L_c \)} = -\sum_{i=1}^{N} y_i \log(p_i)
\label{eq:9}
\end{equation}

\begin{equation}
\text{\( L_d \)} = 1 - \frac{2 \sum_{i=1}^{N} y_i \cdot p_i}{\sum_{i=1}^{N} p_i^2 + \sum_{i=1}^{N} y_i^2}
\label{eq:10}
\end{equation}
where \( N \) represents the total number of pixels in the input image. \( y_i \) denotes the ground truth label for pixel \( i \). \( p_i \) is the predicted probability for pixel \( i \) being part of the corresponding organ.

\begin{figure}[ht]
   \begin{center}
   \includegraphics[width=\textwidth, trim=2cm 5.1cm 2cm 3.8cm, clip]{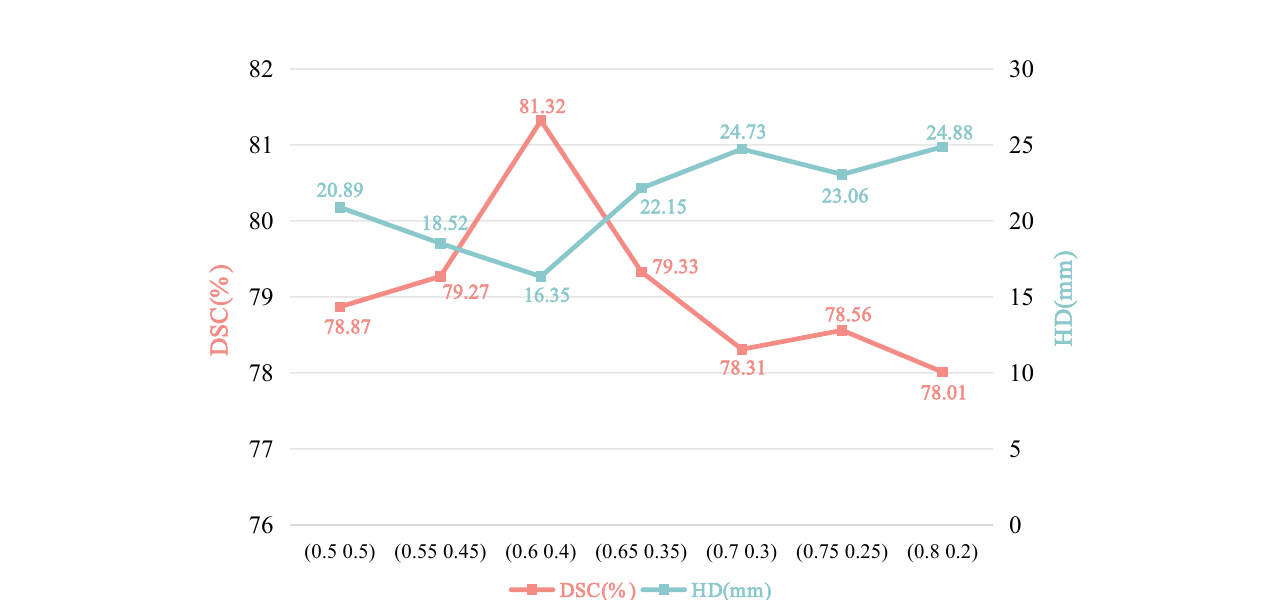} 
   %
   %
   \caption{DSC and HD trends across different loss weighting strategies. The graph illustrates that the DSC (represented by a light red line) peaks and the HD (depicted by a light blue line) reaches its lowest value at the parameter setting of (0.6, 0.4). This alignment confirms it as the optimal parameter setting for segmentation performance.}
   \label{f:5} 
   \end{center}
\end{figure}

To analyze the impact of \( w_c \) and \( w_d \) on the model's performance, an extensive comparison of various combinations of their values is conducted. As shown in Fig.~\ref{f:5}, when \( w_c \) incrementally increases from 0.5 to 0.8 and \( w_d \) simultaneously decreases from 0.5 to 0.2, both DSC and HD exhibit distinct trends. Initially, as the weights change from \( w_c = 0.5, w_d = 0.5 \) to \( w_c = 0.6, w_d = 0.4 \), there is a noticeable improvement in DSC, which peaks at 81.32\%, while HD decreases to 16.35 mm. This trend suggests that enhancing the influence of \( L_c \) contributes positively to the segmentation accuracy in overlapping regions and improves boundary delineation. However, as \( w_c \) further increases to 0.7 and 0.8 with corresponding \( w_d \) values of 0.3 and 0.2, DSC declines to 78.01\%, and HD rises. This reversal indicates a decrease in the model's robustness in boundary segmentation. At \( w_c = 0.5, w_d = 0.5 \), both DSC and HD remain moderate but fail to achieve optimal performance. Overall, the weight combination of \( w_c=0.6, w_d=0.4 \) achieves the best trade-off between segmentation accuracy and boundary robustness. With this setting, the model achieves a DSC of 81.32\% and HD of 16.35 mm, outperforming other weight configurations. This result highlights the importance of balancing \( L_c \) and \( L_d \) to improve the performance of medical image segmentation.

\subsection{Complementary roles of MEWB and DA+ Modules}

To evaluate the effectiveness of the MEWB and DA+ blocks in FMD-TransUNet, we conducted ablation studies by integrating each module individually into the baseline TransUNet architecture. Specifically, the Only MEWB model denotes the baseline model augmented with the MEWB module and the Only DA+ model indicates the integration of DA+ block into the baseline. The detailed integration of these modules within the network framework is elaborated in Section 3.2. 

As summarized in Table~\ref{tab:2}, the baseline TransUNet model achieves an average DSC of 77.48\% and an average HD of 31.69~mm. The Only MEWB model improved the average DSC to \( 79.10\% \), representing a 1.62\% gain over the baseline, while reducing the average HD by 41.8\% to \( 18.45 \, \text{mm} \). Furthermore, the MEWB module demonstrates a clear improvement in DSC for the majority of organs. For example, in segmenting the gallbladder, a small organ, MEWB elevates DSC from 63.13\% to 69.20\%, marking a 9.6\% improvement. Similar advantages are observed in larger organs. The spleen attains a DSC of 88.94\%, reflecting a 3.86\% gain over the baseline, while the liver achieves a DSC of 94.55\%, with a marginal yet consistent improvement of 0.47\%. When the DA+ block is added alone, the model achieves an average DSC of \( 80.28\% \) and an HD of \( 25.14 \, \text{mm} \). Compared to TransUNet, DA+ also improved the DSC for challenging organs, such as the left kidney, where the DSC increases from 81.87\% to 83.50\%, and the stomach, with the DSC rising from 75.62\% to 82.32\%. The observed improvements in DSC and HD metrics demonstrate that both the MEWB and DA+ modules are effective enhancements to the baseline TransUNet.

\begin{table*}[htbp]
\begin{center}
\caption{Ablation study on the segmentation performance of different modules in FMD-TransUNet}
\renewcommand{\arraystretch}{2} 
\setlength{\tabcolsep}{5pt} 

\begin{adjustbox}{max width=\textwidth} 
\label{tab:2}
\begin{tabular}{lcccccccccccccccccc}
\hline
\textbf{Model} & \multicolumn{2}{c}{Average} & \multicolumn{2}{c}{Aorta} & \multicolumn{2}{c}{Gallbladder} & \multicolumn{2}{c}{Kidney (L)} & \multicolumn{2}{c}{Kidney (R)} & \multicolumn{2}{c}{Liver} & \multicolumn{2}{c}{Pancreas} & \multicolumn{2}{c}{Spleen} & \multicolumn{2}{c}{Stomach} \\
\cmidrule(lr){2-3} \cmidrule(lr){4-5} \cmidrule(lr){6-7} \cmidrule(lr){8-9} \cmidrule(lr){10-11} \cmidrule(lr){12-13} \cmidrule(lr){14-15} \cmidrule(lr){16-17} \cmidrule(lr){18-19}
 & DSC & HD & DSC & HD & DSC & HD & DSC & HD & DSC & HD & DSC & HD & DSC & HD & DSC & HD & DSC & HD \\
\hline
TransUNet      &77.48 & 31.69& 87.23 & -     & 63.13 & -     & 81.87 & -     & 77.02 & -     & 94.08 & -     & 55.86 & -     & 85.08 & -     & 75.62 & -  \\
Only MEWB      & 79.10 & 18.45 & 86.85 & \textbf{4.63} & \textbf{69.20} & 27.59 & 80.64 & 23.66 & 76.81 & \textbf{17.61} & 94.55 & 15.16 & 59.69 & 16.20 & 88.94 & 24.67 & 76.10 & \textbf{18.05} \\
Only DA+       & 80.28 & 25.14 & 87.15 & 14.27 & 63.52 & 40.15 & 83.50 & 48.33 & 80.23 & 23.55 & \textbf{94.87} & 14.88 & 61.42 & \textbf{11.52} & 89.25 & 24.32 & \textbf{82.32} & 24.15 \\
FMD-TransUNet  & \textbf{81.32} & \textbf{16.35} & \textbf{88.76} & 20.08 & 65.23 & \textbf{14.99} & \textbf{85.12} & \textbf{5.26}  & \textbf{82.12} & 21.42 & 94.19 & \textbf{12.98} & \textbf{66.31} & 13.91 & \textbf{89.73} & \textbf{23.18} & 79.13 & 18.98 \\
\hline
\end{tabular}
\end{adjustbox}
\end{center}
\vspace{-2em} 
\begin{flushleft}
\fontsize{9}{10}\selectfont {The bold font indicates the best}
\end{flushleft}
\end{table*}

In addition, the MEWB and DA+ modules display distinct yet complementary strengths in the multi-organ segmentation. The Only MEWB model exhibits superior boundary refinement, achieving lower HD than DA+ for most organs. For instance, in segmenting the gallbladder, a small organ, MEWB reduces HD to 27.59 mm, outperforming DA+ by 31.3\%, which attains a HD of 40.15~mm. Similar advantages are observed for the left kidney, where MEWB achieves HD of 23.66 mm versus DA+'s 48.33 mm, and for the stomach, with HD of 18.05 mm compared to DA+'s 24.15 mm. These results suggest that the MEWB module has a distinct advantage in boundary refinement, due to its ability to capture frequency features. In contrast, the DA+ model excels in regional feature enhancement, achieving higher DSC than the MEWB model across seven organs. Specific improvements include: 6.22\% for the stomach, 2.86\% for the left kidney, 1.73\% for the pancreas, 3.42\% for the right kidney, 0.32\% for the liver, 0.31\% for the spleen, and 0.3\% for the aorta. These improvements highlight the strength of DA+ in resolving ambiguities within heterogeneous regions through dynamic attention mechanisms that amplify discriminative local features.

By integrating both modules, the proposed FMD-TransUNet combines their complementary capabilities. It achieves the best overall performance with an average DSC of \( 81.32\% \) and an HD of \( 16.35 \, \text{mm} \), surpassing both single-module configurations. Compared to TransUNet, the DSC improved by \( 3.84\% \), and the HD decreased by \( 15.34 \, \text{mm} \) on average. This configuration markedly improved both segmentation accuracy and boundary precision across most organs. For example, the DSC of the left kidney reached \( 85.12\% \), and its HD dropped to \( 5.26 \, \text{mm} \). Similarly, the spleen, pancreas, and aorta achieved higher DSC compared to single-module models. However, for some organs, the DSC was slightly lower than that of single-module configurations. This phenomenon can be attributed to increased model complexity. 

\begin{figure}[ht]
   \begin{center}
   \includegraphics[width=0.8\textwidth, trim=0cm 0.1cm 4cm 0.2cm, clip]{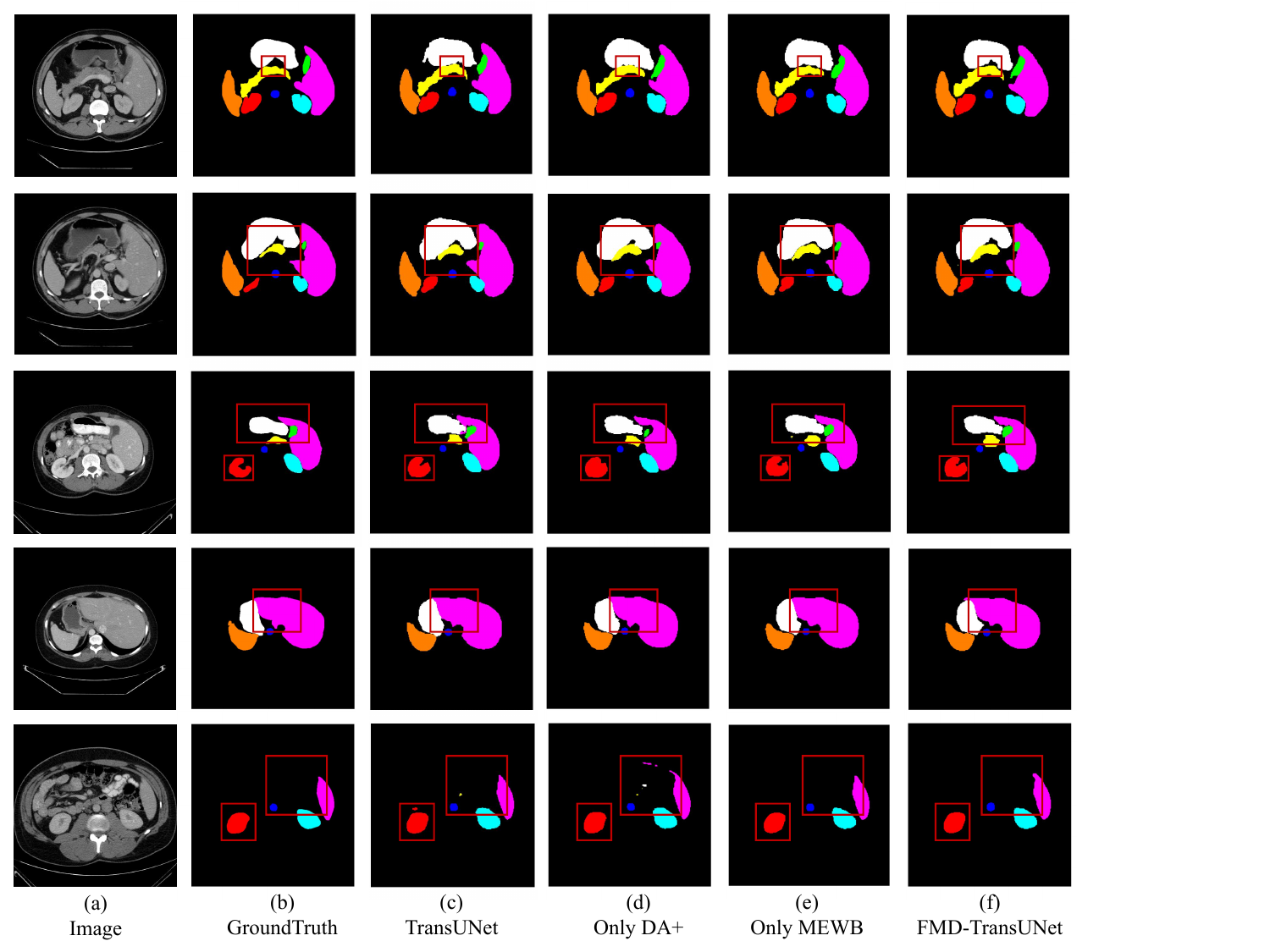} 
   %
   %
   \caption{Qualitative comparison of segmentation results using different modules. From left to right: (a) Image, (b) GroundTruth, (c) TransUNet, (d) Only DA+, (e) Only MEWB, (f) FMD-TransUNet. The red rectangles identify organ regions where the superiority of our proposed method can be clearly seen.}
   \label{f:6} 
   \end{center}
\end{figure}

The qualitative comparison results as shown in Fig.~\ref{f:6} further validate the aforementioned findings. The DA+ block alone improves segmentation for large organs like the liver and spleen, highlighting its strength in regional feature extraction. However, for smaller organs like the gallbladder and kidneys, as shown in the third row, the DA+ block struggles to achieve precise segmentation, leading to boundary errors and inaccuracies. The MEWB module, on the other hand, exhibits notable advantages in boundary refinement. It performs particularly well in segmenting smaller and more complex organs such as the gallbladder, as demonstrated in the third row where the gallbladder is segmented more accurately. However, it has limitations in capturing fine details in small regions such as the pancreas, as can be seen in the first row. The FMD-TransUNet, integrating both MEWB and DA+ blocks, achieves the most accurate and consistent segmentation across all organs. For example, in the fifth row, the pancreas and liver boundaries are accurately segmented, while in the second row, the gallbladder segmentation is better than the single-module models. Additionally, in the third row, the kidney segmentation shows improved detail and precision, maintaining accurate boundaries and shape information.

In conclusion, the qualitative and quantitative analyses jointly show that the MEWB and DA+ modules are highly complementary. The MEWB module primarily contributes global boundary refinement and enhancing HD, while the DA+ module focuses on increasing regional accuracy and improving the DSC score. Their integration achieves an optimal balance between global and local feature extraction, resulting in superior segmentation performance for most organs compared to both the baseline and single-module models. These findings confirm the effectiveness of integrating these modules into the TransUNet framework for precise multi-organ segmentation.

\subsection{Location and numbers of DA+ blocks in skipping connections}

To investigate the influence of DA+ blocks on skip connections at different layers, we performed a series of ablation experiments, as shown in Table~\ref{tab:3}. Without the DA+ block, the model achieves a DSC of 78.59\% and HD of 24.64 mm, which represents the baseline performance. When the DA+ block is added to the first connection, the DSC improves to 79.43\%, and the HD decreases to 21.00 mm, indicating that incorporating DA+ blocks at shallow layers begins to enhance segmentation accuracy and boundary delineation. Extending DA+ to the first and second skip connections further elevates DSC to 80.02\% and reduces HD to 17.62 mm, signifying that intermediate-layer attention mechanisms strengthen global-local feature fusion. Finally, incorporating DA+ across all three skip connections achieves optimal performance with DSC of 81.32\% and HD of 16.35~mm, highlighting the necessity of hierarchical feature refinement. 

This progressive improvement stems from DA+’s dual attention mechanism, which systematically refines spatial and channel dependencies at multiple scales. In the shallow layers, DA+ removes noisy low-level features like texture noise. In deeper layers, it boosts important anatomical feature such as organ boundaries. The cumulative effect ensures that skip connections bridge the semantic gap between encoder and decoder while preserving accuracy. These results validate that DA+ blocks are most effective when integrated across all skip connection layers, ultimately leading to a notable improvement in both regional accuracy and boundary precision.

\begin{table*}[htbp]
\begin{center}
\caption{Ablation study on Adding DA+ Blocks to skip connections in different layers}
\label{tab:3}
\begin{tabular}{lcccccc}
\hline
         & \multicolumn{3}{c}{Skip Connections Added} & \multicolumn{2}{c}{Metrics} \\
\cmidrule(lr){2-4} \cmidrule(lr){5-6}
\textbf & 1 st layer & 2 st layer & 3 st layer & DSC (\%)$\uparrow$ & HD (mm)$\downarrow$ \\
\hline
DA+ block &               &               &               & 78.59          & 24.64         \\
          &               &               &               &                &               \\
DA+ block & \checkmark    &               &               & 79.43          & 21.00         \\
          &               &               &               &                &               \\
DA+ block & \checkmark    & \checkmark    &               & 80.02          & 17.62         \\
          &               &               &               &                &               \\
DA+ block & \checkmark    & \checkmark    & \checkmark    & \textbf{81.32} & \textbf{16.35} \\
\hline
\end{tabular}
\end{center}
\begin{flushleft}
\fontsize{9}{10}\selectfont {The bold font indicates the best}
\end{flushleft}
\end{table*}


\section{Conclusions}

This study presents FMD-TransUNet, an innovative framework for abdominal multi-organ segmentation. By combining the frequency-domain multi-axis representation learning based on MEWB with the enhanced dual attention mechanism based on DA+, our model effectively addresses the limitations of existing approaches. Specifically, the MEWB effectively captures global and local frequency characteristics to complement spatial-domain features, while the DA+ module enhances spatial and channel attention to reduce the semantic gap, jointly improving feature representation. 

Evaluated on the Synapse dataset, FMD-TransUNet outperforms state-of-the-art methods, achieving an average DSC of 81.32\% and HD of 16.35~mm. Ablation studies confirm the complementary roles of MEWB and DA+ in enhancing segmentation accuracy and boundary precision. However, the current model is optimized only for single-modality CT images. Future research will explore the integration of multi-modal data to further enhance segmentation performance.

\bmhead{Funding}
\addcontentsline{toc}{section}{\numberline{}Appendix}
This work is supported by the National Natural Science Foundation of China (Grant No. 11801511) and Zhejiang Provincial Natural Science Foundation of China (LY22A010003).

\bmhead{Data availability}
The original data used is a public dataset, which can be accessed at \url{https://www.synapse.org/#!Synapse:syn3193805/wiki/217789}.

\section*{Declarations}

\bmhead{Conflict of interest} The authors declare that they have no conficts of interest.
\bmhead{Ethics approval} This article does not contain any studies with human participants or animals performed by any of the authors.




\begin{thebibliography}{00}

\bibitem{ref1} Lyu P, Wang C, Zhu J, et al. Application of artificial intelligence algorithms for medical image multi-organ segmentation in the field of medicine. \textit{Progress in Pharmaceutical Sciences}. 2023;47(10):751-757. \url{https://link.cnki.net/doi/10.20053/j.issn1001-5094.2023.10.005}

\bibitem{ref2} Ronneberger O, Fischer P, Brox T. U-Net: Convolutional networks for biomedical image segmentation. In: Proceedings of the International Conference on Medical Image Computing and Computer-Assisted Intervention. Springer; 2015:234-241. \url{https://doi.org/10.1007/978-3-319-24574-4_28}

\bibitem{ref3} Zhou Z, Siddiquee MMR, Tajbakhsh N, et al. UNet++: Redesigning skip connections to exploit multiscale features in image segmentation. \textit{IEEE Transactions on Medical Imaging}. 2020;39(6):1856-1867. 

\bibitem{ref4} Oktay O, Schlemper J, Folgoc LL, et al. Attention U-Net: Learning where to look for the pancreas. \textit{arXiv preprint arXiv:1804.03999}, 2018. \url{https://doi.org/10.48550/arXiv.1804.03999}

\bibitem{ref5} Diakogiannis FI, Waldner F, Caccetta P, et al. ResUNet-a: A deep learning framework for semantic segmentation of remotely sensed data. \textit{ISPRS Journal of Photogrammetry and Remote Sensing}. 2020;162:94-114. \url{https://doi.org/10.1016/j.isprsjprs.2020.01.013}

\bibitem{ref6} Çiçek Ö, Abdulkadir A, Lienkamp SS, et al. 3D U-Net: Learning dense volumetric segmentation from sparse annotation. In: Proceedings of the International Conference on Medical Image Computing and Computer-Assisted Intervention. Springer; 2016:424-432. \url{https://doi.org/10.1007/978-3-319-46723-8_49}

\bibitem{ref7} Jégou S, Drozdzal M, Vazquez D, et al. The One Hundred Layers Tiramisu: Fully convolutional DenseNets for semantic segmentation. In: \textit{Proceedings of the IEEE Conference on Computer Vision and Pattern Recognition Workshops}. 2017:11-19.

\bibitem{ref8} Gu Z, Cheng J, Fu H, et al. CE-Net: Context Encoder Network for 2D Medical Image Segmentation. \textit{IEEE Transactions on Medical Imaging}. 2019;38(10):2281-2292.

\bibitem{ref9} Vaswani A, Shazeer N, Parmar N, et al. Attention is all you need. In: \textit{Advances in Neural Information Processing Systems}. 2017;30.

\bibitem{ref10} Dosovitskiy A, Beyer L, Kolesnikov A, et al. An image is worth 16x16 words: Transformers for image recognition at scale. \textit{arXiv preprint arXiv:2010.11929}, 2020.

\bibitem{ref11} Chen J, Lu Y, Yu Q, et al. TransUNet: Transformers make strong encoders for medical image segmentation. \textit{arXiv preprint arXiv:2102.04306}, 2021. \url{https://doi.org/10.48550/arXiv.2102.04306}

\bibitem{ref12} Wang H, Cao P, Wang J, et al. Uctransnet: Rethinking the skip connections in U-Net from a channel-wise perspective with transformer. In: \textit{Proceedings of the AAAI Conference on Artificial Intelligence}. 2022;36(3):2441-2449. \url{https://doi.org/10.1609/aaai.v36i3.20144}

\bibitem{ref13} Cao H, Wang Y, Chen J, et al. Swin-Unet: Unet-like pure transformer for medical image segmentation. In: \textit{European Conference on Computer Vision}. Cham: Springer Nature Switzerland; 2022:205-218. \url{https://doi.org/10.1007/978-3-031-25066-8_9}

\bibitem{ref14} Sun G, Pan Y, Kong W, et al. DA-TransUNet: Integrating spatial and channel dual attention with transformer U-net for medical image segmentation. \textit{Frontiers in Bioengineering and Biotechnology}. 2024;12:1398237. \url{https://doi.org/10.3389/fbioe.2024.1398237}

\bibitem{ref15} Rao Y, Zhao W, Liu B, et al. Global Filter Networks for Image Classification. In: \textit{Proceedings of the Advances in Neural Information Processing Systems}. 2021;34:980-993.

\bibitem{ref16} Li P, Zhou R, He J, et al. A global-frequency-domain network for medical image segmentation. \textit{Computers in Biology and Medicine}. 2023;164:107290. \url{https://doi.org/10.1016/j.compbiomed.2023.107290}

\bibitem{ref17} Zhang X, Li Z, Xiao B, et al. Multi-axis frequency domain residual UNet for pore identification in deep shale scanning electron microscopy images. In: \textit{2024 International Joint Conference on Neural Networks}. IEEE; 2024: 1-8.


\bibitem{ref18} Xiang H, Xu R, Ma J. HM-ViT: Hetero-modal vehicle-to-vehicle cooperative perception with vision transformer. In: \textit{Proceedings of the IEEE/CVF International Conference on Computer Vision}. 2023:284-295.

\bibitem{ref19} Gehrig M, Scaramuzza D. Recurrent vision transformers for object detection with event cameras. In: \textit{Proceedings of the IEEE/CVF Conference on Computer Vision and Pattern Recognition}. 2023:13884-13893.

\bibitem{ref20} Li X, Ding H, Yuan H, et al. Transformer-based visual segmentation: A survey. \textit{IEEE Transactions on Pattern Analysis and Machine Intelligence}. 2024.

\bibitem{ref21} Zhou HY, Guo J, Zhang Y, et al. nnFormer: Volumetric medical image segmentation via a 3D transformer. \textit{IEEE Transactions on Image Processing}. 2023;32:4036-4045.

\bibitem{ref22} Azad R, Al-Antary M T, Heidari M, et al. TransNorm: Transformer provides a strong spatial normalization mechanism for a deep segmentation model. \textit{IEEE Access}. 2022;10:108205–108215.

\bibitem{ref23} Zhang S, Xu Y, Wu Z, et al. CTC-Net: A Novel Coupled Feature-Enhanced Transformer and Inverted Convolution Network for Medical Image Segmentation. In: \textit{Asian Conference on Pattern Recognition}. Cham: Springer Nature Switzerland; 2023:273-283. \url{https://doi.org/10.1007/978-3-031-47637-2_21}

\bibitem{ref24} Heidari M, Kazerouni A, Soltany M, et al. Hiformer: Hierarchical multi-scale representations using transformers for medical image segmentation. In: \textit{Proceedings of the IEEE/CVF Winter Conference on Applications of Computer Vision}. 2023:6202-6212.

\bibitem{ref25} Rao Y, Zhao W, Zhu Z, et al. Global filter networks for image classification. In: \textit{Advances in Neural Information Processing Systems}. 2021;34:980-993.

\bibitem{ref26} Chen Y, Zhang X, Peng L, et al. Medical image segmentation network based on multi-scale frequency domain filter. \textit{Neural Networks}. 2024;175:106280. \url{https://doi.org/10.1016/j.neunet.2024.106280}

\bibitem{ref27} Tang S, Ran H, Yang S, et al. A frequency selection network for medical image segmentation. \textit{Heliyon}. 2024;10(16). \url{10.1016/j.heliyon.2024.e35698 External Link}

\bibitem{ref28} Liu S, Lin Y, Liu D, et al. Frequency-Enhanced Lightweight Vision Mamba Network for Medical Image Segmentation. \textit{IEEE Transactions on Instrumentation and Measurement}. 2025.

\bibitem{ref29} Fu J, Liu J, Tian H, et al. Dual attention network for scene segmentation. In: \textit{Proceedings of the IEEE/CVF Conference on Computer Vision and Pattern Recognition}. IEEE; 2019: 3146-3154.

\bibitem{ref30} Jha D, Riegler MA, Johansen D, et al. Doubleu-net: A deep convolutional neural network for medical image segmentation. In: \textit{2020 IEEE 33rd International Symposium on Computer-Based Medical Systems}. IEEE; 2020:558-564.

\bibitem{ref31} Lewis J, Cha YJ, Kim J. Dual encoder–decoder-based deep polyp segmentation network for colonoscopy images. \textit{Scientific Reports}. 2023;13(1):1183. \url{https://doi.org/10.1038/s41598-023-28530-2}

\bibitem{ref32} Ding M, Xiao B, Codella N, et al. Davit: Dual attention vision transformers. In: \textit{European Conference on Computer Vision}. Cham: Springer Nature Switzerland; 2022:74-92. \url{https://doi.org/10.1007/978-3-031-20053-3_5}

\bibitem{ref33} Ates GC, Mohan P, Celik E. Dual cross-attention for medical image segmentation. \textit{Engineering Applications of Artificial Intelligence}. 2023;126:107139. \url{https://doi.org/10.1016/j.engappai.2023.107139}

\bibitem{ref34} Wu Y, He K. Group normalization. In: \textit{Proceedings of the European Conference on Computer Vision}. 2018:3–19.

\bibitem{ref40} Sandler M, Howard A, Zhu M, et al. MobileNetV2: Inverted residuals and linear bottlenecks. In: \textit{Proceedings of the IEEE Conference on Computer Vision and Pattern Recognition}. 2018: 4510–4520.

\bibitem{ref36} Fu J, Liu J, Tian H, et al. Dual attention network for scene segmentation. In: \textit{Proceedings of the IEEE/CVF Conference on Computer Vision and Pattern Recognition}. 2019: 3146–3154.

\bibitem{ref35} Chollet F. Xception: Deep learning with depthwise separable convolutions. In: \textit{Proceedings of the IEEE Conference on Computer Vision and Pattern Recognition}. 2017:1251–1258.

\bibitem{ref37} Drozdzal M, Vorontsov E, Chartrand G, et al. The importance of skip connections in biomedical image segmentation. \textit{International Workshop on Deep Learning in Medical Image Analysis}. Cham: Springer International Publishing; 2016:179-187. \url{https://doi.org/10.1007/978-3-319-46976-8_19}

\bibitem{ref38} Landman B, Xu Z, Igelsias J, et al. Miccai multi-atlas labeling beyond the cranial vault–workshop and challenge. In: \textit{Proc. MICCAI Multi-Atlas Labeling Beyond Cranial Vault—Workshop Challenge}. 2015;5:12.

\bibitem{ref39} Huang Y. ViT-R50 GAN: Vision Transformers Hybrid Model based Generative Adversarial Networks for Image Generation. In: \textit{2023 3rd International Conference on Consumer Electronics and Computer Engineering}. 2023: 590–593.

\end{thebibliography}
\end{document}